\documentclass[twocolumn,english,aps,prb,floatfix]{revtex4}
\usepackage[T1]{fontenc}
\usepackage[latin1]{inputenc}
\usepackage{graphicx}
\usepackage{times}

\makeatletter

\usepackage{bm}

\makeatother
\begin{document}

\title{Conductance of deformable molecules with interaction }

\author{J. Mravlje$^{1}$, A. Ram\v{s}ak$^{1,2}$, and T. Rejec$^{1,2,3}$}

\affiliation{$^{1}$J. Stefan Institute, Ljubljana, Slovenia}

\affiliation{$^{2}$Faculty of Mathematics and Physics, University of Ljubljana,
Slovenia}

\affiliation{$^{3}$Department of Physics, Ben-Gurion University, Beer-Sheva,
Israel}

\date{April 18, 2005}
\begin{abstract}
Zero temperature linear response conductance of molecules with Coulomb
interaction and with various types of phonon modes is analysed
together with local occupation, local moment, charge fluctuations and
fluctuations of molecular deformation. Deformation fluctuations are
quantitatively related to charge fluctuations which exhibit similarity
also to static charge susceptibility.
\end{abstract}


\maketitle
The evidence for phonon assisted tunneling was found already in early
conductance measurements in double-barrier heterostructures.\cite{goldman87}
In conductance measurements of nanoscale systems such as quantum dots
or real molecules, the Coulomb interaction leads to the Coulomb blockade
or the Kondo effects.\cite{park02} The electron-phonon interaction
also proved to play an essential role in such systems. In particular,
in single organic molecules electronic transport is influenced by
vibrational fine structure.\cite{park00,zhitenev02}

Phonon degrees of freedom lead to a mass enhancement of a single electron
in the empty conduction band. The problem is known as the conventional
polaron problem.\cite{holstein59} The local form of the polaron problem
arises when the coupling between electrons and phonons is confined
to only one site.\cite{hewson74} Here, the main effect of phonons
is a narrowing of the level width, analogous to the electron-phonon
mass enhancement and is similar to the level width renormalization
due to electron-hole pairs.\cite{haldane78} Theoretical investigations
of the combined effect, the electron-electron and the electron-phonon
interaction, show that quite diverse impurity systems can be described
by Anderson model with renormalized effective parameters.\cite{hewson04}

Early studies of conductance of various types of quantum systems with
electron-phonon interaction were based on the calculation of the transmission
probability as a function of incident energy for an electron interacting
with Einstein phonons as it tunnels through a double-barrier structure.\cite{wingreen89}
The transmission probability for single injected electrons exhibits
phonon-assisted transmission resonances -- side-bands -- at energies
near the main elastic resonance.\cite{wingreen89,bonca95} Such side-bands
appear also in the linear conductance calculation results if the coupling
to the Fermi sea in the leads is not correctly taken into account.\cite{zhu03}
However, vibrational side-bands would be discernible in non-linear
conductance measurements.\cite{bing04,paaske04} Recently, the numerical
renormalization group method applied to a single-molecule device\cite{cornaglia04,cornaglia2}
showed that the problem can in certain regimes be mapped onto the
anisotropic Kondo model.\cite{costi98} Phonon effects in molecular
transistors were investigated also in quantal and classical treatment.\cite{mitra04}

\begin{figure}
\begin{center}\includegraphics[%
  width=60mm,
  keepaspectratio]{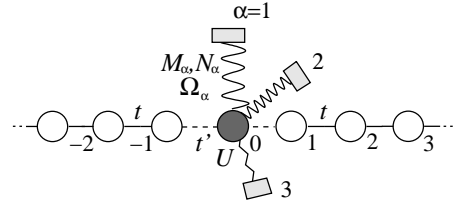}\end{center}

\caption{\label{cap:Fig1}Non-interacting leads attached to an Anderson site,
e.g., molecule (gray-shaded site) with various phonon modes.}
\end{figure}

In this paper we concentrate on the deformation of a molecule in a
linear response conductance measurement. In particular, the molecule
is attached to the left and right non-interacting lead, schematically
presented in Fig.~\ref{cap:Fig1}, and described with the Anderson
model

\begin{equation}
H_{\mathrm{e}}=-\sum_{\left\langle ij\right\rangle \sigma}t_{ij}c_{i\sigma}^{\dagger}c_{j\sigma}+\epsilon_{d}n_{d}+Un_{d\uparrow}n_{d\downarrow},\label{eq:he}\end{equation}
 where $\left\langle ij\right\rangle $ represents nearest neighbor
hopping. In the leads $t_{ij}\equiv t$ and $t_{\pm1,0}=t_{0,\pm1}\equiv t'$
is the hopping matrix element from the leads to the molecule. The
occupation of the molecule is $n_{d}=\sum_{\sigma}n_{d\sigma}$ with
$n_{d\sigma}=d_{\sigma}^{\dagger}d_{\sigma}$, where $d_{\sigma}\equiv c_{0\sigma}$.
The deformation of the molecule is modeled with a general form of
electron-phonon coupling and the molecule coupled to the leads is
described with

\begin{eqnarray}
H & = & H_{\mathrm{e}}+\sum_{\alpha}\Omega_{\alpha}a_{\alpha}^{\dagger}a_{\alpha}+\sum_{\alpha}\biggl[M_{\alpha}\left(n_{d}-1\right)\nonumber \\
 &  & +N_{\alpha}\sum_{\sigma}\left(d_{\sigma}^{\dagger}c_{-1\sigma}+d_{\sigma}^{\dagger}c_{1\sigma}+\mathrm{h.c.}\right)\biggr]x_{\alpha},\label{eq:heph}\end{eqnarray}
 where $M_{\alpha}$ and $N_{\alpha}$ are the local and nearest neighbor
electron-phonon coupling constants corresponding to arbitrary number
of orthogonal vibrational modes with energies $\Omega_{\alpha}$ and
corresponding displacements $x_{\alpha}=a_{\alpha}^{\dagger}+a_{\alpha}$.

The zero temperature linear response conductance through the molecule
is calculated from the sine formula,\cite{rr03,rrfon} $G=G_{0}\sin^{2}[(E_{+}-E_{-})/4tL]$,
where $G_{0}=2e^{2}/h$ and $E_{\pm}$ are the ground state energies
of a large auxiliary ring consisted of $L$ non-interacting sites
and an embedded interacting system (molecule), with periodic and antiperiodic
boundary conditions, respectively. The chemical potential is set at
the middle of the band in the leads, which corresponds to $L$ electrons
in the system. To determine the ground state energy, we generalized
the projection-operator method proposed by Gunnarsson and Sch\"{ o}nhammer.\cite{gunnarsson85}
The variational expression for the ground state is

\begin{equation}
|\Psi\rangle=\sum_{\lambda\left\{ \! m_{\alpha}\!\right\} }C_{\lambda\left\{ \! m_{\alpha}\!\right\} }P_{\lambda}\prod_{\alpha}a_{\alpha}^{\dagger m_{\alpha}}\left|\tilde{0}\right\rangle ,\label{eq:psi}\end{equation}
 where $P_{\lambda}$ are projection operators to multi-electron states
of an isolated molecule, $P_{0}=\left(1-n_{d\uparrow}\right)\left(1-n_{d\downarrow}\right)$,
$P_{1}=\sum_{\sigma}n_{d\sigma}\left(1-n_{d\bar{\sigma}}\right)$,
and $P_{2}=n_{d\uparrow}n_{d\downarrow}$, as well as additional operators
involving hopping of electrons between the molecule and leads (for
details, see Ref.~\onlinecite{gunnarsson85}). The vacuum state $\left|\tilde{0}\right\rangle $
is the ground-state of a decoupled, non-interacting electron-phonon
system, described by renormalized matrix elements $\tilde{t}^{\prime}$
and $\tilde{\epsilon}_{d}$. An approximation to the ground-state
energy is obtained by minimizing the total energy with respect to
coefficients $C_{\lambda\left\{ \! m_{\alpha}\!\right\} }$ and parameters
$\tilde{t}^{\prime}$ and $\tilde{\epsilon}_{d}$ while allowing a
sufficiently large number of excited phonons, in order to obtain a
converged result. 

\begin{figure}
\begin{center}\includegraphics[%
  width=75mm]{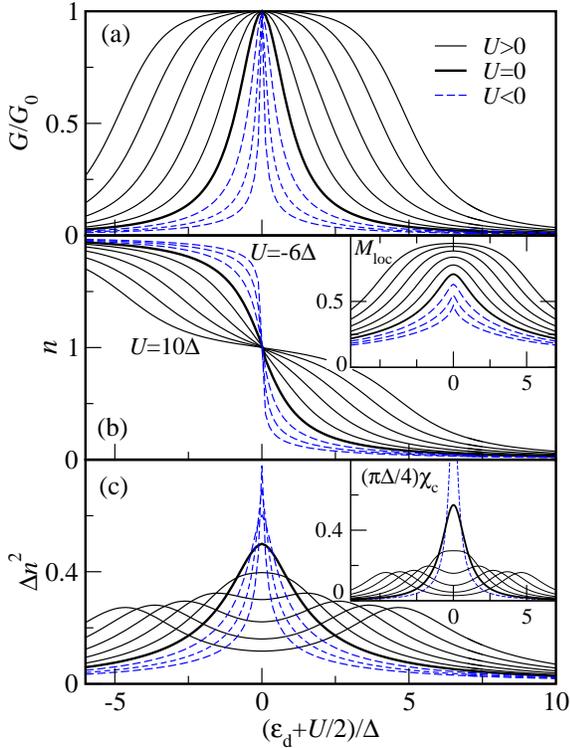}\end{center}

\caption{\label{cap:Fig2} (Color online) (a) Conductance for the bare Anderson
model with $-6\Delta\leq U\leq10\Delta$ in increments of $2\Delta$
(full lines for $U>0$, dashed lines for $U<0$ and a thicker full
line for $U=0$). (b) Local occupancy $n$ and local moment $M_{\mathrm{{loc}}}$
(inset). (c) Charge fluctuations $\Delta n^{2}=2n-n^{2}-M_{\mathrm{{loc}}}^{2}$.
Inset: renormalized charge susceptibility $(\pi\Delta/4)\chi_{c}$.}
\end{figure}

In the limit of large frequencies, $\Omega_{\alpha}$, finite $M_{\alpha}$
and $N_{\alpha}=0$, the model is equivalent to the bare Anderson
model with renormalized parameters $U_{\mathrm{eff}}=U-\sum_{\alpha}2M_{\alpha}^{2}/\Omega_{\alpha}$
and $\epsilon_{d,\mathrm{eff}}=\epsilon_{d}-\sum_{\alpha}M_{\alpha}^{2}/\Omega_{\alpha}$.\cite{hewson74}
We first test our numerical formalism in this regime. In Fig.~\ref{cap:Fig2}(a)
the conductance through an undeformable molecule ($M_{\alpha}=N_{\alpha}=0$),
for various $U$ and a fixed $\Delta=2t^{\prime2}/t=t/5$ is presented. 

For a positive $U$ the conductance exhibits a plateau in the Kondo
regime and the results obtained with the present method accurately
reproduce\cite{rr03} the exact solution based on the Bethe ansatz.\cite{wiegman80}
Local electron density $n=\left\langle n_{d}\right\rangle $ is related
to the conductance through the Friedel sum rule\cite{friedel} and
is characterized with a plateau $n\sim1$ in the Kondo regime, as
presented in Fig.~\ref{cap:Fig2}(b). Kondo physics is signalled
also with the increase of the local moment $M_{\mathrm{{loc}}}=\bigl<\left(n_{d\uparrow}-n_{d\downarrow}\right)^{2}\bigr>^{1/2}$,
presented in the inset. The related occupancy (charge) fluctuations,
$\Delta n^{2}=\bigl<\left(n_{d}-n\right)^{2}\bigr>$, are presented
in Fig.~\ref{cap:Fig2}(c) and in the inset is given the corresponding
charge susceptibility, $\chi_{c}=-\partial n/\partial\epsilon_{d}$.
In accordance with the fluctuation-dissipation theorem, charge fluctuations
are similar to the charge susceptibility, $\Delta n^{2}\sim(\pi\Delta/4)\chi_{c}$.\cite{fdt}
Fluctuations are suppressed in the Kondo regime, and are larger in
the mixed valence regime, $\left|\epsilon_{d}\right|\lesssim\Delta$
or $\left|\epsilon_{d}+U\right|\lesssim\Delta$.

For sufficiently strong electron-phonon coupling $M_{\alpha}$ the
effective electron-electron interaction is attractive, $U_{\mathrm{eff}}<0$.
In this regime the situation is opposite to the more standard spin
Kondo regime because at the impurity the system favors electron (hole)
pairs rather than local moments due to single electrons.\cite{sch84}
Therefore strong charge fluctuations emerge in the particle hole symmetric
point when the chemical potential is level with the local bipolaron
energy leading to charge-fluctuation (anisotropic) Kondo effect.\cite{costi98}
In Fig.~\ref{cap:Fig2}(a) the conductance for various attractive
$U<0$ in the bare Anderson model is presented with dashed lines.
The first observation is a narrowing of the conductance curve and
the corresponding enhanced charge fluctuations {[}Fig.~\ref{cap:Fig2}(c){]},
consistent with a sharp transition in the local occupation and a suppresion
of the local moment, Fig.~\ref{cap:Fig2}(b). For increasing (negative)
$U$, charge susceptibility is not limited and overshoots charge fluctuations.
However, the comparison of $\Delta n^{2}$ and $\chi_{c}$ confirms
at least qualitative proportionality. Our analysis showed that these
results for the renormalized bare Anderson model represent also generic
behavior of the general model with $N_{\alpha}=0$.

\begin{figure}
\begin{center}\includegraphics[%
  width=90mm]{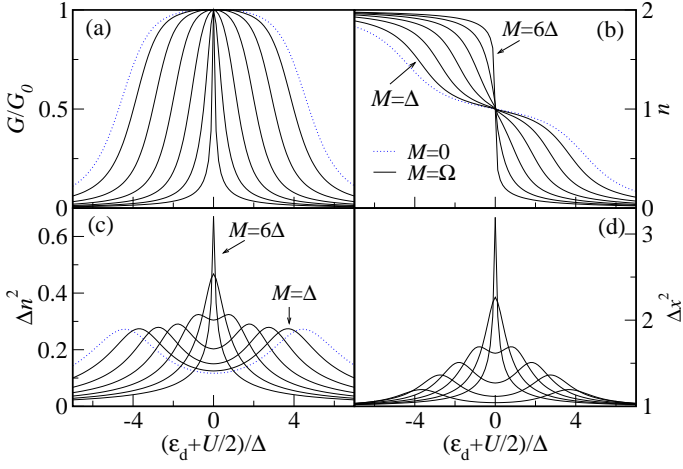}\end{center}

\caption{\label{cap:Fig3} (Color online) Results for $U=10\Delta$ and a
single phonon mode with a fixed $M=\Omega$. (a) Conductance vs. $\left(\epsilon_{d}+U/2\right)/\Delta$
for $0\leq M\leq6\Delta$ in increments of $\Delta$. The dotted line
represents the results for an undeformable molecule, $M=0$. (b) Occupation,
(c) occupation fluctuations and (d) deformation fluctuations. }
\end{figure}

In Fig.~\ref{cap:Fig3} the results of the analysis of a molecule
with a single vibrational mode using $U=10\Delta$ are presented.
The coupling--frequency ratio is kept constant, $M_{\alpha}/\Omega_{\alpha}\equiv M/\Omega=1$,
while the electron-phonon coupling $M$ is varied from $M=\Delta$
to $M=6\Delta$ in increments of $\Delta$. The results for conductance,
occupancy and occupancy fluctuations {[}Figs.~\ref{cap:Fig3}(a,
b, c){]} resemble the results of the bare Anderson model with renormalized
parameters.\cite{hewson04} There are no discernible side-bands in
the conductance. For comparison also the result for undeformable molecule
$(M=0)$ is presented. The molecule deformations, i.e. $x_{\alpha}$,
are related to the occupation of the impurity. The displacement of
individual modes is in general given with $\langle x_{\alpha}\rangle=
-2M_{\alpha}/\Omega_{\alpha}(n-1)-16N_{\alpha}/\Omega_{\alpha}\textrm{Re}
\langle d_{\sigma}^{\dagger}c_{1\sigma}\rangle$.\cite{cornaglia2}
Fluctuations of the deformation, $\Delta x_{\alpha}^{2}=\bigl<\left(x_{\alpha}-\left\langle x_{\alpha}\right\rangle \right)^{2}\bigr>$,
are related to the average number of particular phonons in the system,
$\Delta x_{\alpha}^{2}=1+2\left\langle a_{\alpha}^{\dagger}a_{\alpha}\right\rangle +2\textrm{Re}\left\langle a_{\alpha}^{\dagger}a_{\alpha}^{\dagger}\right\rangle -\left\langle x_{\alpha}\right\rangle ^{2}$.
A deviation of $\Delta x_{\alpha}^{2}$ from unity signals deviations
from the coherent state of the oscillator. In the limit of large phonon
frequencies (fast modes) the oscillator deformations can follow charge
fluctuations, $\Delta x_{\alpha}^{2}-1=(2M_{\alpha}/\Omega_{\alpha})^{2}\Delta n^{2}$,
while in general they are smaller than that. In the limit of small
phonon frequencies (slow modes), phonons feel the average occupation
of the molecule and $\Delta x_{\alpha}^{2}-1=0$. The fluctuations
corresponding to a single mode system, $\Delta x_{\alpha}^{2}\equiv\Delta x^{2}$,
for the same set of parameters and labeled as above, are presented
in Fig.~\ref{cap:Fig3}(d). 

\begin{figure}
\begin{center}\includegraphics[%
  width=70mm]{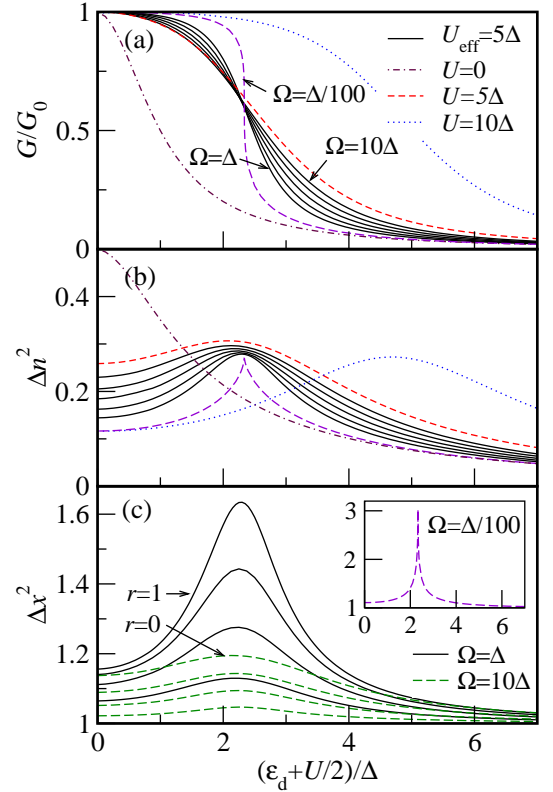}\end{center}

\caption{\label{cap:Fig4} (Color online) A fixed $U_{\mathrm{eff}}=5\Delta$
with $U=10\Delta$ and $\Omega_{1}=\Delta$, $\Omega_{2}=10\Delta$
for various $M_{1,2}$ (corresponding to $r=0$, 1/4, 1/2, 3/4 and
1) - full lines. Also plotted are the results for a bare Anderson
model with $U=10\Delta$, $U=5\Delta$ and $U=0$ (dotted, short-dashed
and dashed-dotted, respectively). Long-dashed lines correspond to
a single softer mode with $\Omega=\Delta/100$ and the same $U_{\mathrm{eff}}=5\Delta$.
(a) Conductance, (b) occupation fluctuations and (c) deformation fluctuations
for modes 1 and 2. In the inset, the deformation fluctuations for
a single softer mode are shown.}
\end{figure}

Next we turn our attention to the case of a molecule with multiple
vibrational modes. Here we give results for the case $U=10\Delta$
with mode 1 with frequency $\Omega_{1}=\Delta$ and mode 2 with frequency
$\Omega_{2}=10\Delta$. The effective Coulomb interaction is reduced
due to coupling to both modes $U_{\mathrm{eff}}=U-2M_{1}^{2}/\Omega_{1}-2M_{2}^{2}/\Omega_{2}$.
We take $N_{\alpha}=0$ and thus the system retains the particle-hole
symmetry. Therefore, in Fig.~\ref{cap:Fig4} only $\epsilon_{d}+U/2>0$
regime is shown. In order to study both, particular and mutual influence
of different modes, we fix $U_{\mathrm{eff}}\equiv5\Delta$, and set
$2M_{1}^{2}/\Omega_{1}=r\left(U-U_{\mathrm{eff}}\right)$, $2M_{2}^{2}/\Omega_{2}=\left(1-r\right)\left(U-U_{\mathrm{eff}}\right)$
where by varying $r$ one can explore the effect of particular modes.
For a single stiffer mode 2, $r=0$, the conductance curve, Fig.~\ref{cap:Fig4}(a),
is suppressed in the Kondo regime and enhanced in the empty orbital
region compared to the softer mode 1, $r=1$. As a limiting case of
this regime the bare Anderson model results for $U=5\Delta$ are presented.

As a representative of the opposite limit, we consider very soft phonons
with $\Omega=\Delta/100$. In the Kondo regime the conductance is
close to the unrenormalized Anderson model result with $U=10\Delta$.
In the mixed valence regime the curve is much steeper, due to strong
renormalization of hopping parameter $\tilde{t}^{\prime}$. In the
empty-orbital regime the conductance approaches the result obtained
with doubly reduced electron-electron interaction $\tilde{U}=U-4M^{2}/\Omega$,
which can be understood as follows. First the oscillator displacement
is shifted, $x\rightarrow\tilde{x}+2\lambda$ thus the Hamiltonian
is transformed into $\tilde{H}=\left(\epsilon_{d}+2\lambda M\right)n_{d}+\tilde{x}\left[M\left(n_{d}-1\right)+\Omega\lambda\right]+\Omega\tilde{a}^{\dagger}\tilde{a}+...$,
where $\lambda=-M\left(n-1\right)/\Omega$, with vanishing transformed
displacement. This Hamiltonian can be solved with trial wave functions
with $m_{\alpha}=0$. Renormalized local energies are then $\epsilon_{d}+2M^{2}/\Omega$,
$\epsilon_{d}$, and $\epsilon_{d}-2M^{2}/\Omega$ for $n=0,1,2$,
respectively. The shifts of $\epsilon_{d}$ where $n=0,\,2$ in turn
correspond to reduced $\tilde{U}=U-4M^{2}/\Omega$ and to $\tilde{U}=U$
for $n=1$.

Occupancy $n$ is related to the conductance similarly as in the previous
figures and is not presented for this case. Charge fluctuations, Fig.~\ref{cap:Fig4}(b),
are similar to the relation between $\Delta n^{2}$ and $\chi_{c}$
as in the above single mode case: the fluctuations are larger for
stiffer phonon modes, except in the mixed-valence regime, where $\Delta n^{2}$
is very weakly dependent of $\Omega_{\alpha}$. Charge fluctuations
for the case of softer mode $\Omega=\Delta/100$ exhibit limiting
behavior consistent with $G$ as discussed above. Deformation fluctuations,
$\Delta x_{\alpha=1,2}^{2}$, are shown in Fig.~\ref{cap:Fig4}(c).
As expected, the fluctuations of the softer mode 1 are enhanced in
comparison with the stiffer mode 2. This effect is even more pronounced
for $\Omega=\Delta/100$ {[}inset of Fig.~\ref{cap:Fig4}(d){]}.

\begin{figure}
\begin{center}\includegraphics[%
  width=65mm]{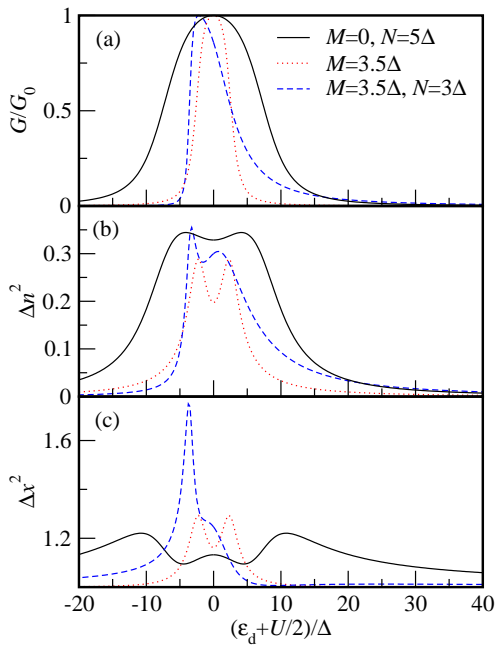}\end{center}

\caption{\label{cap:Fig5} (Color online) (a) Conductance, (b) occupation
fluctuations, and (c) deformation fluctuations for various types of
electron-phonon interaction ($U=10\Delta$, $\Omega=5\Delta$).}
\end{figure}

In Fig.~\ref{cap:Fig5} we present the results for a general case
of single electron-phonon mode coupling with $U=10\Delta$ and $\Omega=5\Delta$.
As pointed out by Cornaglia \emph{et al.}\cite{cornaglia2}, $N_{\alpha}$
terms together with $M_{\alpha}$ break the particle-hole symmetry,
while the symmetry is conserved if only one of the terms is non-vanishing.
In Fig.~\ref{cap:Fig5}(a) the conductance for three typical cases
is shown. Full line represents the $M=0$ and $N=5\Delta$ results
where the resonance width is severely increased, because the $N$-terms
increase the effective $t'$. If both electron-phonon coupling terms
are relevant, e.g., $M=3.5\Delta$ and $N=3\Delta$, the conductance
exhibits asymmetry (dashed line), compared to the $N=0$ case (dotted
line). In Fig.~\ref{cap:Fig5}(b) the corresponding occupation fluctuations
are presented. Displacement fluctuations $\Delta x_{\alpha}$, Fig.~\ref{cap:Fig5}(c),
in this case are not related solely to occupation fluctuations, but
also to fluctuations of the hopping operator $d_{\sigma}^{\dagger}c_{1\sigma}$
(not shown here).

We have presented results of a comprehensive analysis of linear response
conductance through a deformable molecule with electron-electron interaction
and different orthogonal phonon modes. In general, the conductance
does not exhibit side-bands and is related to the Anderson model with
renormalized parameters for the single- or multiple-phonon interaction.
Additionally, the emphasis of our analysis was on the deformation
fluctuations of the molecule due to multiple phonon modes. Phonons
in slow phonon modes are permanently in a coherent state corresponding
to the average occupation of the molecule and the deformation fluctuations
are minimal in this limit (except close to the charge fluctuations
maxima). In the opposite limit of a fast phonon mode, phonons form
a coherent state corresponding to the occupation at a given moment
in time. Therefore, deformation fluctuations are enhanced proportionally
to charge fluctuations, the proportionality coefficient being $-(2M_{\alpha}/\Omega_{\alpha})^{2}$.
In general, the deformation fluctuations take a value between these
two limits. It was also shown that charge fluctuations are approximately
proportional to static charge susceptibility of the molecule. The
method used here proved to be robust and appropriate for a wide range
of generalizations due to specific electron-phonon interaction or
the topology of the interaction region, for example, a molecule with
several interconnected sites.

One of the authors (A.R.) would like to thank A. Hewson for stimulating
discussions. The authors acknowledge the support of the MSZS under
grant Pl-0044.


\begin{thebibliography}{10}
\bibitem{rrfon}In Ref.~\onlinecite{rr03} a general proof that the method is correct
in the case of the electron-electron interaction is given. The main
assumption there is the existence of the Fermi liquid properties of
the interacting system coupled to the leads. For the case of the electron-phonon
coupling studied here the main steps of the proof can be repeated
and the conclussions remain unchanged.
\bibitem{fdt}Strictly, $\langle n_{d}^{2}\rangle$ is given with the integral of
imaginary part of dynamic charge susceptibility, $\chi_{c}^{''}(\omega)$,
therefore the relation to static $\chi_{c}$ is only qualitative (even
at $U=0$). We should note that for large $U$ limit we found reasonable
agreement with $1/{\cal N}$-expansion result $\chi_{c}\sim\frac{\pi}{{\cal N}\Delta}n^{2}(1-n).$
\end{thebibliography}
\end{document}